\documentclass[aps,twocolumn,showpacs]{revtex4}
\usepackage[dvips]{graphics}
\usepackage{color}
\definecolor{gold}{rgb}{0.85,0.66,0}
\definecolor{dblue}{rgb}{0,0,0.8}

\begin{document}

\title{{\textcolor{gold}{Topological effect on spin transport in a 
magnetic quantum wire: Green's function approach}}}

\author{{\textcolor{dblue}{Moumita Dey$^1$, Santanu K. Maiti$^{1,2}$ 
and S. N. Karmakar$^1$}}}

\affiliation{$^1$Theoretical Condensed Matter Physics Division,
Saha Institute of Nuclear Physics, 1/AF, Bidhannagar, Kolkata-700 064,
India \\
$^2$Department of Physics, Narasinha Dutt College, 129 Belilious Road,
Howrah-711 101, India}

\begin{abstract}
We explore spin dependent transport through a magnetic quantum wire 
which is attached to two non-magnetic metallic electrodes. We adopt a 
simple tight-binding Hamiltonian to describe the model where the quantum 
wire is attached to two semi-infinite one-dimensional non-magnetic 
electrodes. Based on single particle Green's function formalism all the 
calculations are performed numerically which describe two-terminal 
conductance and current-voltage characteristics through the wire. Quite
interestingly we see that, beyond a critical system size probability 
of spin flipping enhances significantly that can be used to design a 
spin flip device. Our numerical study may be helpful in fabricating 
mesoscopic or nano-scale spin devices. 
\end{abstract}


\maketitle

\section{Introduction}

Over the past two decades spin dependent transport in low-dimensional 
systems has emerged as one of the most extensively cultivated area in 
condensed matter physics, due to its potential application in the fields 
of nanoelectronics and nanotechnology~\cite{nano1,nano2}. Introduced in 
$1996$, by S. Wolf, the term `spintronics'~\cite{spintronics1,spintronics2, 
spintronics3} refers to a new branch in physics that deals with control 
and manipulation of electron spin in addition to its charge for storage 
and transfer of information as in conventional electronics. Idea of 
possibility to exploit electron spin in transport phenomena was ignited by 
the discovery of giant magneto-resistance (GMR) effect~\cite{gmr} in Fe/Cr 
magnetic multi-layers in $1980$'s and it holds future promises of 
integrating memory and logic into a single device. Since the discovery of 
GMR based magnetic field sensors, revolutionary development has taken place 
in magnetic data storage applications, device processing techniques, and 
quantum computation. `Spintronics' presents a new paradigm in quantum 
computation with incredible speed up in computational time and much 
reduced complexity in quantum-computing algorithm, using the idea of 
quantum coherence and spin entanglement. The key idea of spintronic 
applications involves three basic steps that are injection of spin through 
interfaces, transmission of spin through matter, and finally detection of 
spin. Having considerably larger spin diffusion length molecules and quantum 
confined nanostructures e.g., quantum dots are ideal candidates to study 
spin dependent transmission. Therefore, from technological as well as 
theoretical point of view study of spin transport in mesoscopic regime 
is of great importance today.

Till date several experimental and theoretical works have been done to 
investigate spin transport phenomena at nanoscale level. In $2004$ 
Rohkinson {\em et al.} prepared a spin filter~\cite{rokhinson} using GaAs 
by atomic force microscopy (AFM) with local anodic oxidation and molecular 
beam epitaxy (MBE) methods. In $2005$, gate field controlled 
magnetoresistance~\cite{schonenberger} in carbon nanotube with metallic 
contacts is observed by Sch{\"o}nenberger {\em et al.} In $2007$, Tombros 
{\em et al.} studied spin transport in single graphene layer at room 
temperature~\cite{tombros}. Various other experiments have also been
performed in this field. Along with such novel experimental works spin 
transport in mesoscopic regime has drawn attention from theoretical point 
of view as well. Present theoretical investigations in this field explores 
various interesting features e.g., spin dependent conductance 
modulation~\cite{condmod}, spin filtering~\cite{filter,san1}, spin 
switching~\cite{switch}, spin detecting mechanisms~\cite{detect}, etc. 
Recently, Shokri {\em et al.} have studied spin dependent transmission 
within the coherent transport regime through magnetic or non-magnetic 
nanostructures e.g., quantum wire, etc., attached to semi-infinite magnetic 
or non-magnetic leads using Transfer matrix method and single particle 
Green's function formalism~\cite{shokri1,shokri2,shokri3,shokri4,shokri5}. 
It is observed that the conductance of such mesoscopic systems depends on 
the spin state of electrons passing through the system and can be 
controlled by applying external magnetic field~\cite{shokri6}. 

The aim of our present work is to study spin dependent transmission through
a magnetic quantum wire. The quantum wire, composed of an array of magnetic 
atoms, is attached symmetrically to two non-magnetic (NM) semi-infinite 
one-dimensional ($1$D) metallic electrodes. A simple tight-binding 
Hamiltonian is used to describe the system where all the calculations are 
done by using single particle Green's function formalism~\cite{san2,san3}. 
With the help of Landauer conductance formula~\cite{datta1,datta2}, spin 
dependent conductance is obtained, and the current-voltage characteristics 
are computed from the Landauer-B\"{u}ttiker formalism~\cite{land1,land2,
land3}. Here, we explore various features of spin dependent transport 
using different orientations of local magnetic moments in a magnetic 
quantum wire. It is interesting to note that, {\em for a specific 
configuration of local magnetic moments as we will describe latter, spin 
flip transmission dominates significantly over pure spin transmission after 
the system size becomes larger than a critical value. This phenomenon can 
be utilized to desgin a tailor made spin flip device.}  

The scheme of the paper is as follows. With a brief introduction (Section I), 
in Section II, we describe the model and theoretical formulation for the 
calculation. Section III explores the significant results which describe
two-terminal conductance and current through the wire and our results
clearly depict the spin flipping action depending on the system size. At 
the end, we conclude our results in Section IV.

\section{Model and synopsis of the theoretical formulation}

\begin{figure}[ht]
{\centering \resizebox*{7.8cm}{4.5cm}{\includegraphics{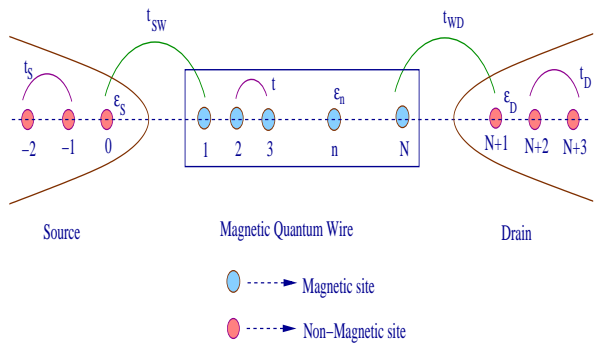}}\par}
\caption{(Color online). A magnetic quantum wire (framed region) of
$N$ atomic sites attached to two semi-infinite one-dimensional
non-magnetic metallic electrodes, namely, source and drain. The filled 
sky blue circles correspond to magnetic atomic sites in the quantum 
wire, while the filled pink circles represent non-magnetic atomic 
sites comprising the electrodes.}
\label{wire}
\end{figure}

We start by describing our model as shown in Fig~\ref{wire}. In this 
figure we illustrate schematically the $1$D nanostructure through which 
we are interested to explore several features of spin dependent transport 
phenomena. In the present work, spin transmission is investigated through 
a magnetic quantum wire (MQW) which is basically an array of $N$ number of 
magnetic atomic sites. Each site has a localized magnetic moment of equal 
amplitude associated with it. The orientations of the local magnetic
moment in a site $n$ (say) are specified by the polar angle $\theta_n$ 
and azimuthal 
angle $\phi_n$ in spherical polar coordinate system and the direction of 
the moment can be altered by applying an external magnetic field. The QW 
is attached symmetrically to two $1$D semi-infinite non-magnetic metallic 
electrodes, commonly termed as source and drain having chemical potentials 
$\mu_1$ and $\mu_2$ under the non-equilibrium condition when bias voltage 
is applied. Described by the discrete lattice model, the electrodes are 
assumed to be composed of infinite non-magnetic sites labeled as $0$, $-1$, 
$-2$, $\ldots$, $-\infty$ for the source and $(N+1)$, $(N+2)$, $(N+3)$, 
$\ldots$, $\infty$ for the drain. 

\begin{figure}[ht]
{\centering \resizebox*{7cm}{3cm}{\includegraphics{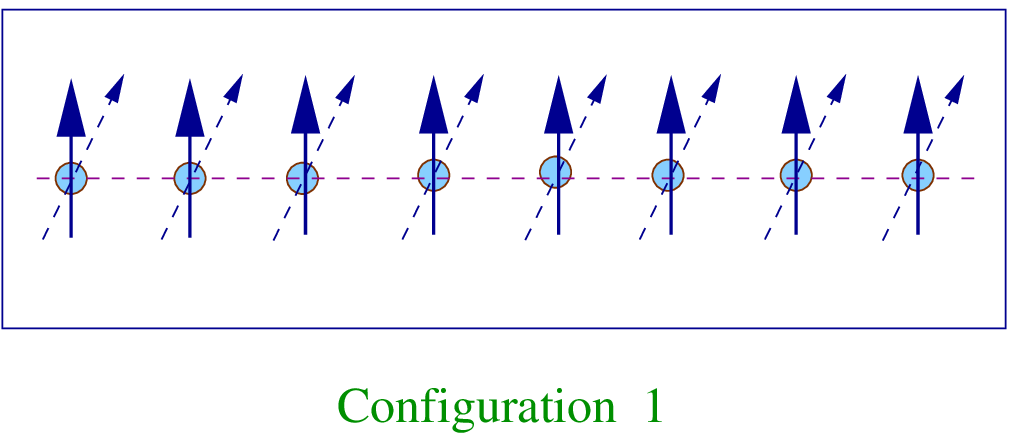}}\par}
\caption{(Color online). Configuration $1$: A magnetic quantum wire with
$N$ atomic sites where all magnetic moments are aligned in a particular
direction (solid arrows). These moments are rotated equally from their 
initial directions by applying an external magnetic field (dashed arrows).}
\label{config1}
\end{figure}

\begin{figure}[ht]
{\centering \resizebox*{7cm}{3cm}{\includegraphics{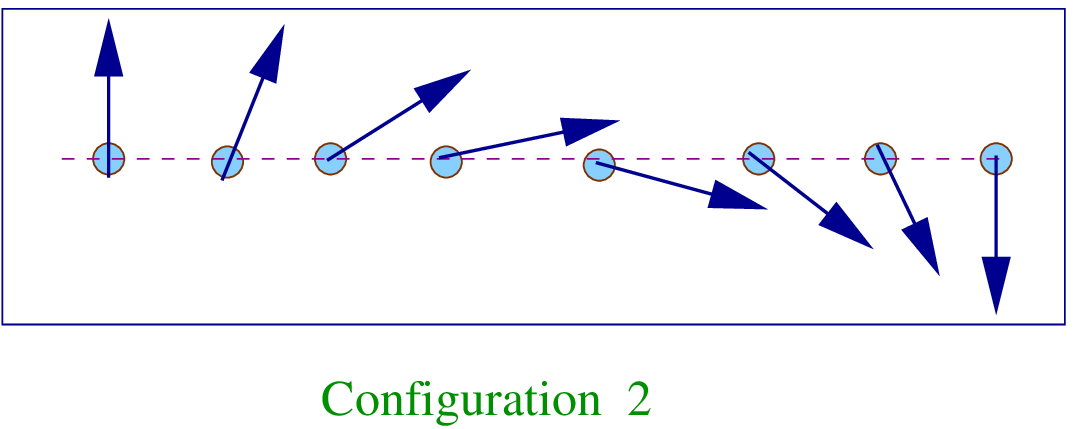}}\par}
\caption{(Color online). Configuration $2$: A magnetic quantum wire with
$N$ atomic sites where the magnetic moments are oriented in a wave like 
pattern. The moments in sites $1$ and $N$ are oppositely oriented.}
\label{config2}
\end{figure}

In our present work we consider two different configurations of the
MQW depending on the orientation of the localized moments on each site
as illustrated in Figs.~\ref{config1} and \ref{config2}. In configuration 
$1$ (see Fig~\ref{config1}), all the magnetic moments are directed along 
$+Z$ direction and their orientation can be changed in an equal amount 
by applying an external magnetic field. In configuration $2$ (see 
Fig~\ref{config2}), the magnetic moments are aligned in a spin wave like 
pattern with the help of a spatially inhomogeneous external magnetic field.    
 
The Hamiltonian for the whole system can be written as,
\begin{equation}
H=H_{W}+H_S+H_D+H_{SW}+H_{WD}
\label{equ1}
\end{equation}
where, $H_{W}$ represents the Hamiltonian for the magnetic quantum wire
(MQW). $H_{S(D)}$ corresponds to the Hamiltonian for the left (right) 
electrode, namely, source (drain), and $H_{SW(WD)}$ is the Hamiltonian 
describing the wire-electrode coupling.

The spin polarized Hamiltonian for the MQW can be written in a single 
electron picture within the framework of tight-binding formulation in 
Wannier basis, using nearest-neighbor approximation as,
\begin{eqnarray}
H_{W} = \sum_{n=1}^N {\bf c_n^{\dagger} \left(\epsilon_0
-\vec{h_n}.\vec{\sigma} \right) c_n} + \sum_{i=1}^N 
{\bf \left(c_i^{\dagger}tc_{i+1} + h.c. \right)}
\label{equ2}
\end{eqnarray}
where, \\
${\bf c_n^{\dagger}}=\left(\begin{array}{cc}
c_{n \uparrow}^{\dagger} & c_{n \downarrow}^{\dagger} \end{array}\right);~~$
${\bf c_n}=\left(\begin{array}{c}
c_{n \uparrow} \\
c_{n \downarrow}\end{array}\right);~~$
${\bf \epsilon_0}=\left(\begin{array}{cc}
\epsilon_0 & 0 \\
0 & \epsilon_0 \end{array}\right)$ \\
${\bf t}=t\left(\begin{array}{cc}
1 & 0 \\
0 & 1 \end{array}\right);~~$ 
${\bf \vec{h_n}.\vec{\sigma}} = h_n\left(\begin{array}{cc}
\cos \theta_n & \sin \theta_n e^{-i \phi_n} \\
\sin \theta_n e^{i \phi_n} & -\cos \theta_n \end{array}\right)$ \\
~\\
\noindent
First term of Eq.~(\ref{equ2}) represents the effective on-site
energies of the atomic sites in the wire. $\epsilon_0$'s are the 
site energies, while the term ${\bf \vec{h_n}.\vec{\sigma}}$ describes
the interaction of the spin (${\bf \sigma}$) of the injected electron
with the localized on-site magnetic moments. On-site flipping of spins
is described mathematically by this term. Second term describes the 
nearest-neighbor hopping between the sites of the quantum wire. 

Similarly, the Hamiltonian $H_{S(D)}$ for the two electrodes can be 
written as,
\begin{equation}
H_{S(D)}=\sum_i {\bf c_i^{\dagger} \epsilon_{S(D)} c_i} + \sum_i
{\bf \left(c_i^{\dagger} t_{S(D)} c_{i+1} + h.c. \right)}
\label{equ3}
\end{equation}
where, $\epsilon_{S(D)}$'s are the site energies of source (drain) and 
$t_{S(D)}$ is the hopping strength between the nearest-neighbor sites 
of source (drain). 
\vskip 0.1cm
\noindent
Here also, \\
~\\
${\bf \epsilon_{S(D)}}=\left(\begin{array}{cc}
\epsilon_{S(D)} & 0 \\
0 & \epsilon_{S(D)} \end{array}\right);~~~
{\bf t_{S(D)}}=\left(\begin{array}{cc}
t_{S(D)} & 0 \\
0 & t_{S(D)} \end{array}\right)$ \\
~\\
\noindent
The wire-electrode coupling Hamiltonian is described by,
\begin{equation}
H_{SW(WD)}= {\bf \left(c_{0(N)}^{\dagger} t_{SW(WD)} c_{1(N+1)} + 
h.c.\right)}
\label{equ4}
\end{equation}
where, $t_{SW(WD)}$ being the wire-electrode coupling strength.

In order to calculate spin dependent transmission probabilities and 
current through the magnetic quantum wire we use single particle 
Green's function technique. Within the regime of coherent transport and 
for non-interacting systems this formalism is well applied.

The single particle Green's function representing the full system for 
an electron with energy $E$ is defined as~\cite{datta1,datta2},
\begin{equation}
\bf{G}=(\bf{E}-\bf{H})^{-1}
\label{equ5}
\end{equation}
where, 
\begin{equation}
{\bf{E}} = (\epsilon + i \eta) {\bf{I}}
\label{equ6}
\end{equation}
In the above expression $i \eta$ is a small imaginary term added to the
energy $\epsilon$ to make the Green's function $\bf{G}$ non-hermitian.

Now, $\bf{H}$ and $\bf{G}$ representing the Hamiltonian and the Green's 
function for the full system those can be partitioned in terms of 
different sub-Hamiltonians like~\cite{datta1,datta2},
\begin{equation}
\bf{H}=\left(\begin{array}{ccc}
\bf{H_{S}} & \bf{H_{SW}} & 0 \\
\bf{H_{SW}^\dag} & \bf{H_{W}} & \bf{H_{WD}}\\
0 & \bf{H_{WD}^\dag} & \bf{H_{D}}\\
\end{array} \right) 
\label{equ7}
\end{equation}
\begin{equation}
\bf{G}=\left(\begin{array}{ccc}
\bf{G_{S}} & \bf{G_{SW}} & 0 \\
\bf{G_{SW}^\dag} & \bf{G_{W}} & \bf{G_{WD}}\\
0 & \bf{G_{WD}^\dag} & \bf{G_{D}}\\
\end{array} \right) 
\label{equ8}
\end{equation}
where, $\bf{H_{S}}$, $\bf{H_{W}}$ and $\bf{H_{D}}$ represent the 
Hamiltonians (in matrix form) for source, quantum wire and drain, 
respectively. $\bf{H_{SW}}$ and $\bf{H_{WD}}$ are the matrices for the 
Hamiltonians representing the wire-electrode coupling. Assuming that 
there is no direct coupling between the electrodes themselves, the 
corner elements of the matrices are zero. A similar definition goes 
for the Green's function matrix ${\bf G}$ as well.

Our first goal is to determine $\bf{G_{W}}$ (Green's function for the 
wire only) which defines all physical quantities of interest. Following 
Eq.~(\ref{equ5}) and using the block matrix form of $\bf{H}$ and $\bf{G}$ 
the form of $\bf{G_{W}}$ can be expressed as, 
\begin{equation}
\bf{G_{W}}=(\bf{E}-\bf{H_{W}}-\bf{\Sigma_{S}}-
\bf{\Sigma_{D}})^{-1}
\label{equ9}
\end{equation}
where, $\bf{\Sigma_{S}}$ and  $\bf{\Sigma_{D}}$ represent the contact 
self-energies introduced to incorporate the effects of semi-infinite 
electrodes coupled to the system, and, they are expressed by the 
relations~\cite{datta1,datta2},
\begin{eqnarray}
\bf{\Sigma_{S}} & = & \bf{H_{SW}^{\dag} G_{S} 
H_{SW}} \nonumber \\
\bf{\Sigma_{D}} & = &  \bf{H_{WD}^{\dag} G_{D} 
H_{WD}}
\end{eqnarray}
Thus, the form of self-energies are independent of the nano-structure 
itself through which transmission is studied and they completely describe 
the influence of electrodes attached to the system. Now, the transmission 
probability $(T_{\sigma \sigma^{\prime}})$ of an electron with energy $E$ 
is related to the Green's function as,
\begin{eqnarray}
T_{\sigma \sigma^{\prime}} & = & {\bf \Gamma}^{1}_{\bf{S}(\sigma \sigma)}
{\bf G}^{1N}_{\bf{r} (\sigma \sigma^{\prime})} 
{\bf G}^{N1}_{\bf{a} (\sigma^{\prime} \sigma)}
{\bf \Gamma}^{N}_{\bf{D}(\sigma^{\prime} \sigma^{\prime})} \nonumber \\
& = & {\bf \Gamma}^{1}_{\bf{S}(\sigma \sigma)}
|{\bf G}^{1N}_{ (\sigma \sigma^{\prime})}|^2
{\bf \Gamma}^{N}_{\bf{D}(\sigma^{\prime} \sigma^{\prime})}
\label{equ11}
\end{eqnarray}
where,
${\bf \Gamma}^{1}_{\bf{S}(\sigma \sigma)} = \langle 1 \sigma| 
{\bf \Gamma_S} | 1 \sigma \rangle $ ,
${\bf \Gamma}^{N}_{\bf{D}(\sigma^{\prime} \sigma^{\prime})} = 
\langle N \sigma^{\prime}| {\bf \Gamma_D} |N \sigma^{\prime} \rangle $ 
and ${\bf G}^{1 N}_{\sigma \sigma^{\prime}} =
\langle 1 \sigma| {\bf G} |N \sigma^{\prime} \rangle $.
%%%%%%%%%%%%%%%%%%%%%%%%%%%%%%%%%%%%%%%%%%%%%%%%%%%%%%%%%%%%%
Here, $\bf{G_{r}}$ and  $\bf{G_{a}}$ are the retarded and advanced single 
particle Green's functions (for the MQW only) for an electron with energy 
$E$. $\bf{\Gamma_{S}}$ and  $\bf{{\Gamma_{D}}}$ are the coupling matrices, 
representing the coupling of the magnetic quantum wire to source and drain, 
respectively, and they are defined by the relation~\cite{datta1,datta2},
\begin{equation}
\bf{\Gamma_{S(D)}} = i[\Sigma^r_{S(D)} - \Sigma^{a}_{S(D)}]
\label{equ12}
\end{equation}
Here, $\bf{\Sigma^r_{S(D)}}$ and $\bf{\Sigma^a_{S(D)}}$ are the retarded 
and advanced self-energies, respectively, and they are conjugate to each 
other. It is shown in literature by Datta {\em et al.} that the self-energy 
can be expressed as a linear combination of a real and an imaginary part 
in the form,
\begin{equation}
{\bf{\Sigma^r_{S(D)}}} = {\bf\Lambda_{S(D)}} - i {\bf\Delta_{S(D)}}
\label{equ13}
\end{equation}
The real part of self-energy describes the shift of the energy levels
and the imaginary part corresponds to the broadening of the levels. The 
finite imaginary part appears due to incorporation of the semi-infinite 
electrodes having continuous energy spectrum. Therefore, the coupling 
matrices can easily be obtained from the self-energy expression and is
expressed as,
\begin{equation}
{\bf{\Gamma_{S(D)}}}=-2 {\bf Im} {\bf{(\Sigma_{S(D)})}}
\label{equ14}
\end{equation}
Considering linear transport regime, conductance $(g_\sigma)$ is obtained
using Landauer formula~\cite{datta1,datta2},
\begin{equation}
g_{\sigma \sigma^{\prime}}=\frac{e^2}{h}T_{\sigma \sigma^{\prime}}
\label{equ15}
\end{equation}
Knowing the transmission probability ($T_{\sigma \sigma^{\prime}}$) of 
an electron injected with spin $\sigma$ and transmitted with spin 
$\sigma^{\prime}$, the current ($I_{\sigma \sigma^{\prime}}$) through the 
system is obtained using Landauer-B\"{u}ttiker formalism. It is written 
in the form~\cite{datta1,datta2},
\begin{equation}
I_{\sigma \sigma^{\prime}} (V)= \frac{e}{h} \int \limits_{-\infty}^{+\infty} 
\left[f_S(E)-f_D(E)\right] T_{\sigma \sigma^{\prime}}(E)~dE
\label{equ16}
\end{equation}
where, $f_{S(D)}=f(E-\mu_{S(D)})$ gives the Fermi distribution function 
of the two electrodes having chemical potentials $\mu_{S(D)}=E_{F} 
\pm eV/2$. $E_F$ is the equilibrium Fermi energy.

\section{Numerical results and discussion}

We investigate various features of spin dependent transport through a
magnetic quantum wire (MQW) for two different geometrical configurations 
depending on the orientations of the localized magnetic moments 
associated with each atomic site. In the first configuration, all the 
moments are aligned parallel to each other and initially they are chosen 
to be directed along $+Z$ direction, while in the second one the magnetic 
moments are oriented in a wave like pattern. All the results shown 
in the present work are obtained numerically. Therefore, we start analyzing 
the results by mentioning all the parameters those are used for numerical 
calculation. Our first assumption is that the two non-magnetic 
side-attached electrodes are made up of identical materials. The on-site 
energies in the wire ($\epsilon_0$) as well as in the leads 
($\epsilon_{S(D)}$) are set as $0$. Hopping strength between the sites in 
the two electrodes is chosen as $t_{S(D)}=4$, whereas in the QW it is set 
as $t=3$. The equilibrium Fermi energy $E_F$ is fixed at $0$. Our unit 
system is simplified by choosing $h=c=e=1$. Energy scale is fixed in unit 
of $t$.                  

Throughout the analysis we address the basic features of spin dependent
transport for two distinct regimes of electrode-to-MQW coupling. These 
regimes are described as follows.
\vskip 0.2cm
\noindent
\underline{Case 1:} Weak-coupling limit
\vskip 0.1cm
\noindent
This limit is set by the criterion $t_{SW(WD)}<<t$. In this case, we 
choose the values as $t_{SW}=t_{WD}=0.5$.

\vskip 0.1cm
\noindent
\underline{Case 2:} Strong-coupling limit
\vskip 0.1cm
\noindent
This limit is described by the condition $t_{SW(WD)} \sim t$. In this
regime we choose the values of hopping strengths as $t_{SW}=t_{WD}=2.5$.

\subsection{Features of Spin Transport for Configuration $1$}

\subsubsection{Conductance-energy characteristics}

\begin{figure}[ht]
{\centering \resizebox*{8cm}{8cm}{\includegraphics{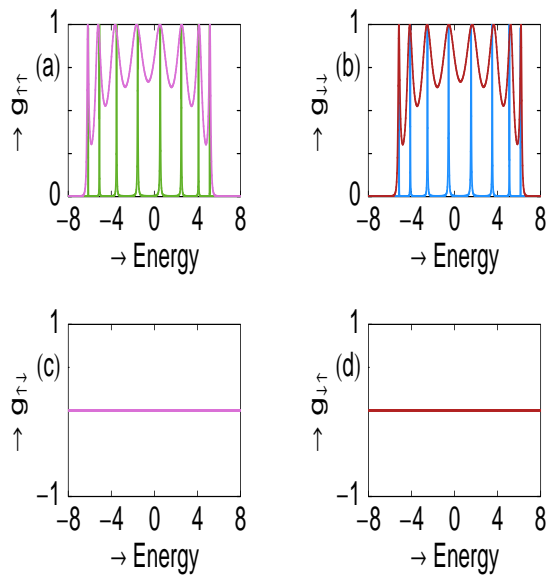}}\par}
\caption{(Color online). $g$-$E$ characteristics for a magnetic quantum 
wire considering the system size $N=8$ where all moments are aligned 
along $+Z$ direction. The upper and lower panels describe the variations 
of $g_{\uparrow \uparrow}$, $g_{\downarrow \downarrow}$ 
and $g_{\uparrow \downarrow}$, $g_{\downarrow \uparrow}$, respectively. 
The green and blue curves represent the results in the weak-coupling 
limit, while the curves in pink and red depict the results in the strong 
coupling limits, respectively. Conductances are measured in unit of 
$e^2/h$, while the energy is measured in unit of $t$.}
\label{cond1}
\end{figure}

First, we plot the conductance-energy characteristics for a magnetic 
quantum wire in which the moments are aligned along a preferred $Z$ 
direction according to the configuration $1$ (for instance see 
Fig.~\ref{config1}). As illustrative examples, in Fig.~\ref{cond1} we 
show the variation of conductances due to pure transmission of up and 
down spin electrons ($g_{\uparrow \uparrow}$ and $g_{\downarrow 
\downarrow}$) and spin flip transmission ($g_{\uparrow \downarrow}$ and 
$g_{\downarrow \uparrow}$) with respect to the injecting electron energy 
$E$ for a magnetic quantum wire considering $N=8$. It is observed that 
$g_{\uparrow \uparrow}$ and $g_{\downarrow \downarrow}$ exhibit sharp 
resonant peaks at some discrete energy values in the weak-coupling limit 
(green and blue curves 
of Figs.~\ref{cond1}(a) and (b)), whereas the peaks acquire some broadening 
in the limit of strong-coupling (pink and red curves of Figs.~\ref{cond1}(a) 
and (b)). The broadening of conductance peaks with the enhancement in 
coupling strength is quantified by the imaginary parts $\bf {\Delta_S}$ and 
$\bf {\Delta_D}$ of the self-energy matrices $\bf{\Sigma_S}$ and 
$\bf{\Sigma_D}$ which are incorporated in the conductance expression via 
the coupling matrices~\cite{datta1,datta2}. 
All these resonant peaks are associated with the energy eigenvalues of 
the magnetic quantum wire. Therefore, it is manifested that the 
conductance-energy spectrum reveals the feature of energy spectrum 
of the wire completely. One interesting feature to be noted here is that,
variation of $g_{\uparrow \uparrow}$ and $g_{\downarrow \downarrow}$ 
with energy ($E$) are exactly mirror symmetric about $E=0$ as we set 
$\epsilon_0=0$. For any other non-zero value of $\epsilon_0$, up and 
down spin channels get splitted but the conductance spectra does not 
remain mirror symmetric anymore.

On the other hand, in this particular configuration the conductance by 
spin flipping becomes exactly zero for the entire energy region as seen 
from Figs.~\ref{cond1}(c) and (d), where the curves for the weak- and 
strong-coupling limits overlap to each other. The reason of zero spin 
flipping is explained as follows. Occurrence of spin flipping is governed 
by the term $\vec{h}.\vec{\sigma}$ in the Hamiltonian (see Eq.~(\ref{equ2})), 
where $\vec{\sigma}$ stands for the Pauli spin matrix having components 
$\sigma_x$, $\sigma_y$ and $\sigma_z$ for the injecting electron and 
$h$ being the localized magnetic moments associated with each magnetic 
site in the quantum wire (QW). Spin flipping is mathematically expressed 
by the operation of raising ($\sigma_+ =\sigma_x + i\sigma_y$) and 
lowering ($\sigma_- =\sigma_x - i\sigma_y$) operators. For the local 
\begin{figure}[ht]
{\centering \resizebox*{8cm}{8cm}{\includegraphics{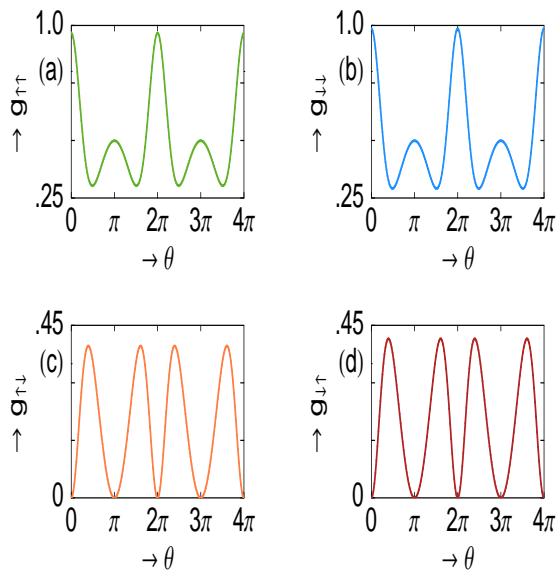}}\par}
\caption{(Color online). Variation of conductances with polar angle $\theta$ 
for a magnetic quantum wire considering $N=16$ and $E=3.2$ in the limit of 
strong wire-electrode coupling strength. The upper and lower panels 
correspond to the cases of $g_{\uparrow \uparrow}$, $g_{\downarrow 
\downarrow}$ and $g_{\uparrow \downarrow}$, $g_{\downarrow \uparrow}$, 
respectively.}
\label{theta}
\end{figure}
magnetic moments oriented along $\pm$ $Z$ axis i.e., for 
$\theta=0~\mbox{and}~\pi $, $\vec{h}.\vec{\sigma}$ ($= h_x\sigma_x + 
h_y\sigma_y + h_z\sigma_z$) becomes equal to $h_z\sigma_z$. Accordingly, 
the Hamiltonian does not contain $\sigma_x$ and $\sigma_y$ and so as 
$\sigma_+$ and $\sigma_-$, which provides zero flipping for up or down 
orientation of magnetic moments. 

\subsubsection{Variation of conductance with polar angle $\theta$}

To reveal the effect of orientation of local magnetic moments on spin 
dependent transport, in Fig.~\ref{theta} we plot the variation of 
conductances ($g_{\uparrow \uparrow}$, $g_{\downarrow \downarrow}$ 
and $g_{\uparrow \downarrow}$, $g_{\downarrow \uparrow}$) with
angle $\theta$ made by the magnetic moments with the preferred $+Z$ 
direction. It is evident from this figure that the conductances 
($g_{\uparrow \uparrow}$, $g_{\downarrow \downarrow}$, 
$g_{\uparrow \downarrow}$ and $g_{\downarrow \uparrow}$) exhibit 
$2\pi$ periodicity as a function of $\theta$ with reflection symmetry 
at $\theta= \pi$. 

Rotation of magnetic moments through an angle $2\pi$ maps themselves into 
their initial positions. So $2\pi$ periodicity in the variation of 
conductances with $\theta$ is expected. Reflection symmetry about 
$\theta = \pi$ is equivalent to having a symmetry point at $\theta=0$, 
which means rotation of magnetic moments from up to down direction is 
independent of the sense of rotation.

\subsubsection{Current-voltage characteristics}

All the essential features of spin transport described earlier will be 
more transparent from our current-voltage characteristics. Current through 
the MQW is computed by integrating over the transmission curve following 
Landauer-B\"{u}ttiker formalism. Transmission probability varies in an 
\begin{figure}[ht]
{\centering \resizebox*{7.8cm}{8cm}{\includegraphics{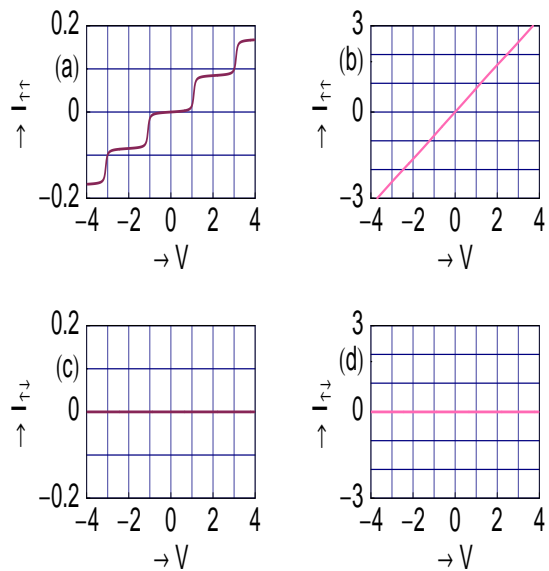}}\par}
\caption{(Color online). $I_{\uparrow \uparrow}$ and 
$I_{\uparrow \downarrow}$ as a function of applied bias voltage $V$ for a 
magnetic quantum wire with $N=8$, where all magnetic moments are aligned 
along $+Z$ direction. The left and right columns correspond to the 
cases of weak and strong wire to electrode coupling limits, respectively.
$I$ and $V$ are measured in units of $te/h$ and $t/e$, respectively.}
\label{current1}
\end{figure}
exactly identical way to that of the conductance spectrum apart from a 
scale factor $e^2/h$ (which is equal to $1$ in our chosen unit system)
according to Eq.~(\ref{equ15}). The current $I_{\uparrow \uparrow}$ shows 
staircase like pattern (Fig.~\ref{current1}(a)) due to the presence of 
sharp, discrete resonant peaks in conductance-energy spectrum in the 
limit of weak-coupling. With the increase in applied bias voltage $V$, the 
difference in chemical potentials of the two electrodes ($\mu_S - \mu_D$) 
increases, allowing more number of energy levels to fall in that range, and 
accordingly, more energy channels are accessible to the injected electrons 
to pass through the magnetic quantum wire from source to drain. 
Incorporation of a single discrete energy level i.e., a discrete quantized 
conduction channel, between the range ($\mu_S - \mu_D$) provides a jump in 
the $I$-$V$ characteristics. Contribution to the current $I_{\uparrow 
\downarrow}$ due to spin flipping is zero (Fig.~\ref{current1}(c)) as the 
spin flip transmission probability is zero for this configuration $1$.

In the limit of strong wire-electrode coupling, due to the broadening 
of conductance peaks current $I_{\uparrow \uparrow}$ shows nearly linear
variation (Fig.~\ref{current1}(b)) as a function of applied bias voltage 
\begin{figure}[ht]
{\centering \resizebox*{8cm}{8cm}{\includegraphics{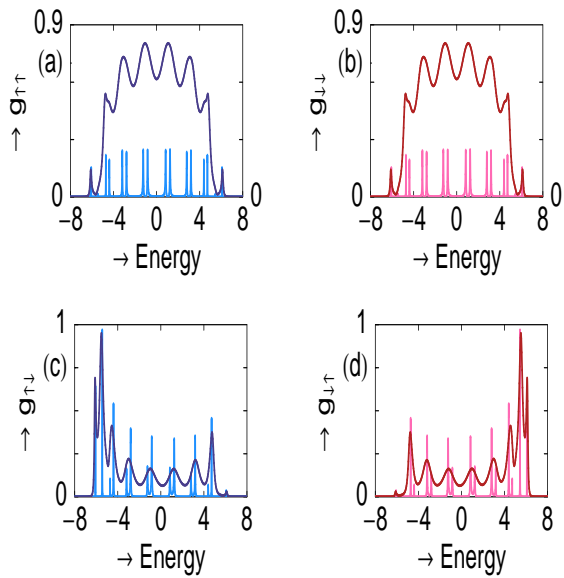}}\par}
\caption{(Color online). $g$-$E$ characteristics for a magnetic quantum 
wire considering the system size $N=8$ where the magnetic moments are
aligned according to configuration $2$ (Fig.~\ref{config2}). The upper 
and lower panels describe the variations of $g_{\uparrow \uparrow}$,
$g_{\downarrow \downarrow}$ and $g_{\uparrow \downarrow}$,
$g_{\downarrow \uparrow}$, respectively. The blue and pink curves 
represent the results in the weak-coupling limit, while the deep blue
and deep red lines correspond to the results in the strong-coupling limit.}
\label{cond2}
\end{figure}
$V$ and acquires much higher amplitude compared to the weak-coupling limit. 
Enhancement in coupling strength does not change spin flip transmission 
probability, and hence, $I_{\uparrow \downarrow}$ shows zero value for 
the entire range of the bias voltage $V$ (Fig.~\ref{current1}(d)).
Current due to down spin shows the same kind of variation with applied 
bias voltage. So this has not been shown in the above figure.

\subsection{Features of Spin Transport for Configuration $2$}

Following the above description of spin dependent transport now we 
concentrate on a magnetic quantum wire in which the moments are 
aligned in a wave like pattern as illustrated in Fig.~\ref{config2}.

\subsubsection{Conductance-energy characteristics}

As representative examples, in Fig.~\ref{cond2} we plot the variation 
of conductances due to pure spin transmission ($g_{\uparrow \uparrow}$
and $g_{\downarrow \downarrow}$) and spin flip transmission 
($g_{\uparrow \downarrow}$ and $g_{\downarrow \uparrow}$) for a MQW 
considering $N=8$. Variation of $g_{\uparrow \uparrow}$ and 
$g_{\downarrow \downarrow}$ exhibits sharp peaks at some discrete 
energy values in the weak-coupling limit (blue and pink curves of 
Figs.~\ref{cond2}(a) and (b)), while in the limit of strong-coupling 
they achieve substantial broadening with larger amplitude 
\begin{figure}[ht]
{\centering \resizebox*{7.8cm}{8cm}{\includegraphics{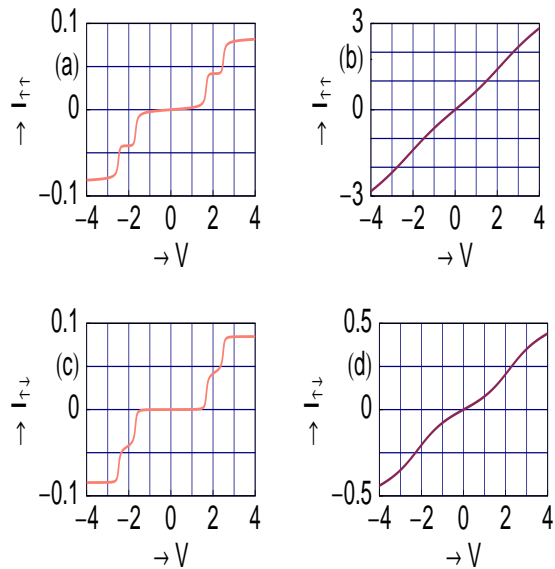}}\par}
\caption{(Color online). $I_{\uparrow \uparrow}$ and 
$I_{\uparrow \downarrow}$ as a function of applied bias voltage $V$ for 
a magnetic quantum wire with $N=8$, where the magnetic moments are
aligned according to the configuration $2$ (Fig.~\ref{config2}). The left 
and right columns correspond to the cases of weak and strong wire to 
electrode coupling strengths, respectively.}
\label{current2}
\end{figure}
(deep blue and deep red curves of Figs.~\ref{cond2}(a) and (b)). This 
broadening effect is clearly understood from our previous discussion. 
But for this type of configuration i.e., where the orientation of each 
magnetic moment is increased gradually with respect to $+Z$ axis 
along the length of the quantum wire, conductance amplitude due to 
pure spin transmission ($g_{\uparrow \uparrow}$ and 
$g_{\downarrow \downarrow}$) in the limit of strong-coupling is 
sufficiently higher than that of the weak-coupling case. 

For this typical configuration, non-zero conductance due to spin flip 
transmission is obtained as given in Figs.~\ref{cond2}(c) and (d). Here, 
the magnitude of spin flip conductances are comparable in both the two
coupling limits. As in the previous case, conductance-energy spectrum due 
to up and down spins are mirror symmetric to each other across the energy
$E=0$, both for pure spin transmission as well as spin flip transmission,
since we set $\epsilon_0=0$ in this case also.

\subsubsection{Current-voltage characteristics}

In Fig.~\ref{current2} we show the variations of spin dependent currents 
$I_{\uparrow \uparrow}$ and $I_{\uparrow \downarrow}$ as a function 
of applied bias voltage $V$ for a magnetic quantum wire considering the
system size $N=8$ in both the two coupling limits. All the basic 
features of currents are same as we observe in the case of configuration 
$1$ e.g., step-like behavior in the weak-coupling limit and almost linear 
variation with larger amplitude in the limit of strong-coupling. But the 
notable signatures for this configuration are non-zero current due to 
spin flipping $I_{\uparrow \downarrow}$, and comparable amplitude of 
$I_{\uparrow \uparrow}$ and $I_{\uparrow \downarrow}$ in the weak-coupling 
limit for this system size. Here also the down spin current shows the same 
nature of variation with the applied bias voltage and correspondingly we 
do not plot it. 

\subsubsection{Variation of conductance with system size $N$}

\begin{figure}[ht]
{\centering \resizebox*{7.8cm}{4.5cm}{\includegraphics{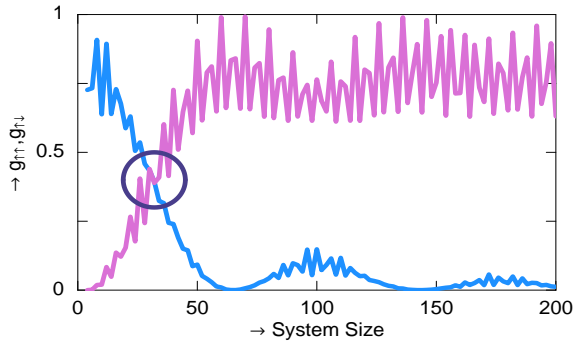}}\par}
\caption{(Color online). Variations of $g_{\uparrow \uparrow}$ and 
$g_{\uparrow \downarrow}$ with system size $N$ for the typical energy 
$E=1.5$ in the limit of strong wire-electrode coupling. The blue and 
pink curves correspond to the features of $g_{\uparrow \uparrow}$ and
$g_{\uparrow \downarrow}$, respectively.}
\label{system1}
\end{figure}

At the end, in Fig.~\ref{system1} we present the variations of $g_{\uparrow 
\uparrow}$ and $g_{\uparrow \downarrow}$ with system size $N$ in the
limit of strong-coupling. The conductances are calculated at the typical
energy $E=1.5$. For such configuration $g_{\uparrow \downarrow}$ increases 
gradually with system size $N$, while $g_{\uparrow \uparrow}$ decreases with
the rise of $N$. For this configuration, the magnetic moments are oriented 
sequentially from $0$ to $\pi$ and from earlier discussion it is evident 
that spin flip does occur when the magnetic moments are oriented at an 
angle $\theta$ with respect to the preferred $+Z$ direction. Hence, for 
the configuration $2$, $g_{\uparrow \downarrow}$ increases with system 
size due to large number of sequential spin flip scattering and after a 
critical system size marked by the violet circle in Fig.~\ref{system1}, 
spin flip transmission dominates significantly over pure spin transmission.
The critical size decreases with increase in injecting electron energy
which is not shown in Fig.~\ref{system1}. In case of down spin propagation 
similar features of system size dependence is observed and we do not show 
this.  

\section{Closing remarks}

In a nutshell, in the present work we study spin dependent transport
through a magnetic quantum wire (MQW) using single particle Green's
function technique. We have adopted a discrete lattice model in 
tight-binding framework to illustrate the system which is simply an 
array of identical magnetic atomic sites. In our theoretical study two 
different geometrical configurations of the MQW depending on the 
orientation of the magnetic moments associated with each magnetic site. 
In the first configuration, all the moments are aligned at an equal 
angle with respect to the $+Z$ axis, while in the second one the 
moments are sequentially oriented from angle $0$ to $\pi$ relative to 
the $+Z$ direction. Orientation of the magnetic moments can be changed 
by applying an external magnetic field.

We investigated conductance-energy ($g$-$E$) and current-voltage ($I$-$V$)
characteristics for both the two configurations mentioned above. Non-zero
spin flip conductances ($g_{\uparrow \downarrow}$ and
$g_{\uparrow \downarrow}$) are obtained for configuration $2$, whereas
it is zero for up or down orientation of localized moments as illustrated
in configuration $1$. In addition to these we observed the dependence of 
conductances on polar angle $\theta$ for configuration $1$, showing $2\pi$
periodicity. In the last figure, we have plotted the variations of 
$g_{\uparrow \uparrow}$ and $g_{\uparrow \downarrow}$ with the system size 
$N$ for typical electron energy $E=1.5$ in the limit of strong wire-electrode 
coupling, which is the most interesting part of our theoretical study. It 
clearly demonstrates that after a certain system size spin flip transmission 
dominates significantly over the pure spin transmission. For a sufficiently 
large system, spin inversion takes place most prominently which can be 
utilized for fabricating spin based nano devices.

In the present work we have calculated all these results by ignoring the 
effects of temperature, spin-orbit interaction, electron-electron 
correlation, electron-phonon interaction, disorder, etc. Here, we set the 
temperature at $0$K, but the basic features will not change significantly 
even in non-zero finite (low) temperature region as long as thermal energy 
($k_BT$) is less than the average energy spacing of the energy levels of 
the magnetic quantum wire. In this model it is also assumed that the two 
side-attached non-magnetic electrodes have negligible resistance.

All these predicted results using such simple geometric configurations 
may be useful in designing a spin based nano devices.


\begin{thebibliography}{99}

\bibitem{nano1} J. Chen, M. A. Reed, A. M. Rawlett, and J. M. Tour,
Science \textbf{286}, 1550 (1999).
\bibitem{nano2} P. Ball, Nature (London) \textbf{404}, 918 (2000).
\bibitem{spintronics1} S. A. Wolf {\em et al.}, Science \textbf{294}, 
1488 (2001).
\bibitem{spintronics2} G. Prinz, Science \textbf{282}, 1660 (1998).
\bibitem{spintronics3} G. Prinz, Phys. Today \textbf{48}, 58 (1995).
\bibitem{gmr} M. N. Baibich, J. M. Broto, A. Fert, F. N. Van Dau, 
F. Petroff, P. Etienne, G. Creuzet, A. Friederich, and J. Chazelas,
Phys. Rev. Lett. \textbf{61}, 2472 (1998).
\bibitem{rokhinson} Rokhinson {\em et al.}, Phys. Rev. Lett.
\textbf{93}, 146601 (2004).
\bibitem{schonenberger} Sch\"{o}nenberger {\em et al.}, Nature Phys. 
\textbf{1}, 99 (2005).
\bibitem{tombros} Tombros {\em et al.}, Nature. \textbf{448}, 571 (2007).
\bibitem{condmod} I. A. Shelykh, N. T. Bagraev, N. G. Galkin, and
L. E. Klyanchkin, Phys. Rev. B \textbf{71}, 113311 (2005).
\bibitem{filter} H. W. Wu, J. Zhou, and Q. W. Shi, Appl. Phys. Lett.
\textbf{85}, 1012 (2004).
\bibitem{san1} M. Dey, S. K. Maiti, and S. N. Karmakar, Phys. Lett. A
\textbf{374}, 1522 (2010).
\bibitem{switch} D. Frustaglia, M. Hentschel, and K. Richter, Phys.
Rev. Lett. \textbf{87}, 256602 (2001).
\bibitem{detect} R. Ionicioiu and I. D'Amico, Phys. Rev. B \textbf{67},
041307(R) (2003).
\bibitem{shokri1} A. A. Shokri, M. Mardaani, and K. Esfarjani, 
Physica E \textbf{27}, 325 (2005).
\bibitem{shokri2} A. A. Shokri, M. Mardaani, and K. Esfarjani, Physica E
\textbf{27}, 325 (2005).
\bibitem{shokri3} A. A. Shokri and M. Mardaani, Solid State Commun. 
\textbf{137}, 53 (2006).
\bibitem{shokri4} M. Mardaani and A. A. Shokri, Chem. Phys. 
\textbf{324}, 541 (2006).
\bibitem{shokri5} A. A. Shokri and A. Daemi, Eur. Phys. J. B
\textbf{69}, 245 (2009).
\bibitem{shokri6} A. A. Shokri and A. Saffarzadeh, J. Phys.: Condens. 
Matter \textbf{16}, 4455 (2004).
\bibitem{san2} S. K. Maiti, Phys. Lett. A \textbf{373}, 4470 (2009).
\bibitem{san3} S. K. Maiti, J. Phys. Soc. Jpn. \textbf{78}, 114602 (2009).
\bibitem{datta1} S. Datta, {\em Electronic transport in mesoscopic systems},
Cambridge University Press, Cambridge (1997).
\bibitem{datta2} S. Datta, {\em Quantum Transport: Atom to Transistor},
Cambridge University Press, Cambridge (2005).
\bibitem{land1} R. Landauer, Phys. Lett. A \textbf{85}, 91 (1986).
\bibitem{land2} R. Landauer, IBM J. Res. Dev. \textbf{32}, 306 (1988).
\bibitem{land3} M. B\"{u}ttiker, IBM J. Res. Dev. \textbf{32}, 317 (1988).
\end{thebibliography}
\end{document}